# Empirical estimation of anthropogenic and natural contributions to surface air temperature trends at different latitudes

*I.I. Mokhov,*[1,2] *D.A. Smirnov*[1,3]

[1] A.M. Obukhov Institute of Atmospheric Physics of the Russian Academy of Sciences, Moscow, Russia, mokhov@ifaran.ru

[2] Lomonosov Moscow State University, Moscow, Russia

[3] Saratov Branch, Kotelnikov Institute of Radioengineering and Electronics of the Russian Academy of Sciences, Saratov, Russia, smirnovda@yandex.ru

How strong are quantitative contributions of the key natural modes of climate variability and the anthropogenic factor characterized by the changes of the radiative forcing of greenhouse gases in the atmosphere to the trends of the surface air temperature at different latitudes of the Northern and Southern Hemispheres on various time intervals? Such contributions to trends are estimated here from observation data with the simplest empirical models. Trivariate autoregressive models are fitted to the data since the 19[th] century and used to assess the impact of the anthropogenic forcing together with different natural climate modes including Atlantic Multidecadal Oscillation, El-Nino / Southern Oscillation, Interdecadal Pacific Oscillation, Pacific Decadal Oscillation, and Antarctic Oscillation. For relatively short intervals of the length of two or three decades, we note considerable contributions of the climate variability modes which are comparable to the contributions of the greenhouse gases and even exceed the latter. For longer intervals of about half a century and greater, the contributions of greenhouse gases dominate at all latitudes as follows from the present analysis of data for polar, middle and tropical regions.

## 1. Introduction

Quantitative estimation of the role of natural and anthropogenic factors in the contemporary climate change is a key problem of the 21[st] century. An overall increase of the global surface air temperature (GST) revealed from the observation data since the 19[th] century includes periods of its faster rise and periods of its decrease. In the beginning of the 21[st] century, the tendency of the global warming slowdown or "hiatus" has arisen. Still, the GST values in the last years are among the highest ones in the entire observational data set since the 19[th] century. With probability greater than 90%, more than a half of the GST rise since the middle of the 20[th] century is attributed to the anthropogenic rise of the atmospheric content of the greenhouse gases (GHGs) [1].

The significant impact of the GHGs (especially $CO_2$) rise on the contemporary GST increase has been revealed from empirical data in many studies which take into account different natural factors including solar and volcanic activity, quasi-cyclic processes like El Nino / Southern Oscillation (ENSO), Atlantic Multidecadal Oscillation (AMO) and others, e.g., [1-32]. In the time range of interest, natural variability essentially enhances or weakens the global warming via producing periods of a faster warming and periods with almost no warming. In particular, Ref. [7] estimates the impacts of the anthropogenic forcing, El Nino phenomena, solar activity, and volcanic activity which altogether explain up to three quarters of the temperature variance since the end of the 19[th] century. Those authors note that along with the dominating role of the anthropogenic factor, El Nino phenomena induce the GST changes up to 0.2 ºC on the time scales of several years, considerable volcanic eruptions – up to 0.3 ºC, and solar activity variations – about 0.1 ºC (see also Ref. [10, 11]). Many other works [2, 3, 8, 11-13, 18, 19, 21, 23, 26, 27] confirm the presence and statistical significance of the GHGs impact on GST and compare it with the impact of other factors using different methods. However, along with such estimates of the impact of different factors on the global climate, it is necessary to obtain concrete numerical estimates of their contributions to the temperature trends for different areas of the Earth from empirical data. In particular, Refs. [29, 30] use a simple method based on trivariate autoregressive (AR) models, which is employed also in this work, and provide estimates of the contributions of GHGs and AMO to the GST trends and the trends of the temperatures at different latitudes of the Northern Hemisphere (NH). This work aims at estimating the contributions of a larger set of significant natural modes of climate variability (AMO, ENSO, Interdecadal Pacific Oscillation – IPO, Pacific Decadal Oscillation – PDO, and Antartic Oscillation – AAO also called Southern Annular Mode) along with GHGs at various latitudes of both NH and Southern Hemisphere (SH). Moreover, we address the question of how strongly the natural climate variability with the time scales up to several decades can enhance or weaken the warming at different latitudes on different time intervals, with a special attention to the intervals of about half a century and shorter. This is principally important, in particular, for the quantitative explanation of the differences between the contemporary temperature trends at various NH and SH latitudes, including polar and subpolar latitudes with very different behaviors of the Arctic and Antarctic sea ice extents [1, 33]. Indeed, apart from empirical estimates, climate models are able to reproduce many modes of climate variability, but not all significant modes of climate variability and their changes are modeled well enough [34].

It is worth to note that the question under study is somewhat similar but still considerably different from detection and attribution of the climate change to one of several selected factors addressed e.g. in Refs. [35-55]. In the detection studies, *a priori* specified climate models (which are usually coupled atmosphere and ocean general circulation models) are used whose multiple realizations are generated under the control experiment condition, i.e. with a constant factor under consideration (e.g. constant atmospheric GHGs content if the impact of this factor is examined). The observed climate fields can be projected onto some typical model patterns called "fingerprints" [35, 37] which are often obtained from model climate change simulations and necessary to enhance detection capability (statistical power, "signal-to-noise" ratio [35, 37]) of the method. If the scale factor (i.e. the corresponding component of the observed fields) significantly differs from the control experiment values, one rejects the hypothesis of no impact of the factor under study whose change is included in the climate change simulations (e.g. a certain scenario of the GHGs rise). In the attribution studies, several factors can be considered in turn with comparison of the observed data with the model fields obtained in different climate change experiments, each including a change of only one factor. If only one of the experiments agrees well with the observed data (after adjusting the scale factor), then the observed climate change is attributed to the impact of that factor. In those studies, one typically assumes that only one factor contributes considerably to the observed climate change. All climate model parameters are not estimated from data and the entire model is supposed to be adequate. The climate field under study is supposed to be a linear combination of internal climate variability and the above patterns corresponding to the climate change experiments, each with a single factor included. Under that setting, it is indeed difficult to perform numerical estimation of simultaneous contributions of different factors to the temperature trends, because it is difficult to carry out multiple climate change experiments with numerous possible combinations of different factors with different weights. One the other hand, if one studies the roles of different factors just in a coupled general circulation model, it is not an empirical estimation since model parameters are not estimated from data.

To avoid the existing difficulties, one can try to assess the contributions to trends from different factors in a simple stochastic dynamical model whose parameters are estimated from the observed data. Probably, it is questionable whether such a model is adequate for the description of climate dynamics under different conditions, but the adequacy of the large general circulation models is also not completely assured. The advantage of a simple model is that it relies just on data and a simple form of the evolution equation without any assumptions concerning parameterizations, etc. Even if the simple empirical models used here appear to be too simple in the future, it is reasonable to start purely empirical numerical estimation of the simultaneous contributions of different factors to the temperature trends with such models.

Below, the data are described in Section 2. Methods are given in Section 3. Results are reported in Section 4. Section 5 concludes.

## 2. Data under analysis

Figure 1 presents the time series under study: interannual variations of the surface air temperature $T$ at different latitudes of the NH and SH (Fig. 1,a) and the indices of AMO, ENSO, IPO, PDO, and AAO with the corresponding GHGs radiative forcing (Fig. 1,b). Temperature anomalies are given relatively to the reference period 1971 – 2000. We note deficiencies of the data for Antarctic latitudes, so the estimates for these latitudes presented below are of a more qualitative than a quantitative character. Selection of the above indices of the natural climate variability is determined by the attempt to include the long-term internal variability associated either with Pacific Ocean (ENSO, IPO, PDO, AAO) or Atlantic Ocean (AMO, NAO) into the empirical AR model for temperature variations. Only one of these indices (except the NAO index) at a time is included into the model along with the GHGs index to keep the model as simple as possible, represent both anthropogenic factor and internal variability, and retain reasonable statistical properties. (Taking into account, along with the AMO index, the NAO index with high interannual variability is not expected to reveal an additional significant contribution to the temperature trends on the intervals of the length of several decades.)

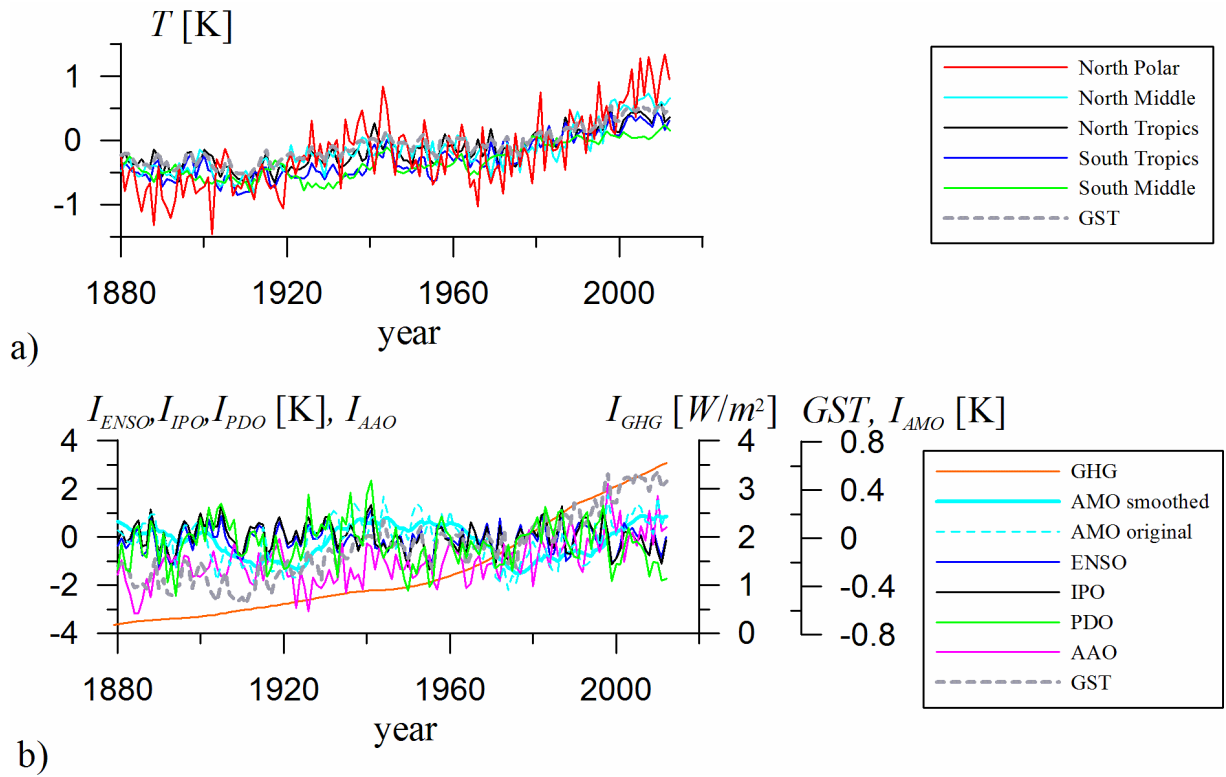

Figure 1. Time series under analysis represent interannual variations of (a) surface air temperature $T$ for the entire Earth (GST, grey dashes) and at various latitudes of the NH and the SH (the legend); (b) indices of AMO (filtered index is shown with cyan solid line, unfiltered one with cyan dashed line), ENSO (blue), IPO (black), PDO (green), and AAO (magenta) along with the GHGs radiative forcing (brown).

To represent the surface air temperatures at various latitudes including the tropical (0–30ºN), middle (30–60ºN), and Arctic (60–90ºN) latitudes of the NH and the corresponding (0–30ºS, 30–60ºS, and 60–90ºS) latitudes of the SH, we have used the mean annual data since 1880 till now (Fig.1,a). The data are the land – ocean temperatures from the Extended Reconstructed Sea Surface Temperature data set (ERSST [56, 57], version 4) located at ftp://ftp.ncdc.noaa.gov/pub/data/noaaglobaltemp/operational/timeseries/ in the following six files (accessed on 08/16/2017): aravg.mon.land.60N.90N.v4.0.1.201706.txt, aravg.mon.land.30N.60N.v4.0.1.201706.txt, aravg.mon.land.00N.30N.v4.0.1.201706.txt,

aravg.mon.land.30S.00N.v4.0.1.201706.txt, aravg.mon.land.60S.30S.v4.0.1.201706.txt, aravg.mon.land.90S.60S.v4.0.1.201706.txt (the data set identifier is doi:10.7289/V5KD1VVF, the detailed description of the data is given at https://www.ncdc.noaa.gov/data-access/marineocean-data/extended-reconstructed-sea-surface-temperature-ersst-v4).

The anthropogenic influences are characterized by the GHGs radiative forcing over 1851 – 2012 with the main contribution from $CO_2$. The forcing data set as used in GISS Climate Models [58] is located at https://data.giss.nasa.gov/modelforce/Miller_et_2014/Fi_Miller_et_al14_upd.txt (accessed on 10/10/2016).

Among the key modes of natural climate variability, we have used the index of AMO since 1856 (HadISST1 data set [59, 60]). Detrended AMO data are located at https://psl.noaa.gov/data/correlation//amon.us.long.data where the index is calculated north of $0^o$. For comparisons to the previous results [29, 30], we have used the previous version of the index computed for the band $20^oN$–$70^oN$ (Fig.1,b, cyan dashed line) located at http://www.esrl.noaa.gov/psd/data/correlation//amon.us.long.data (accessed on 05/04/2017, the detailed description of the data is given at https://psl.noaa.gov/data/timeseries/AMO/). The detrended AMO index possesses the characteristic period of about 6 decades. We have used different filers to focus on a slow component of the AMO index. As a representative case, the main results are shown here for the annual-mean data of the AMO index smoothed with a weighted moving average filter with 10-year window length and weights linearly decreasing with the time lag (i.e. the triangular temporal profile) down to zero at the lag of 11 years as in Ref. [61].

The strongest interannual variability of the global surface air temperature is associated with the El-Nino phenomena. We have used the ENSO index since 1870 till now (HadISST1 data set [60]) which shows sea surface temperature anomalies in the region Nino-3,4 in the equatorial Pacific (5°S–5°N, 170–120°W) and is located at https://psl.noaa.gov/gcos_wgsp/Timeseries/Data/nino34.long.anom.data (accessed on 10/10/2016). Its plot is given in Fig.1,b with the blue line.

We have estimated also the contribution of the PDO and IPO to the surface temperature trends based on the HadISST1.1 data set [60]. We have used the PDO index since 1854 till now based on NOAA's extended reconstruction of sea surface temperatures (ERSST Version 5, (https://www.ncdc.noaa.gov/teleconnections/pdo/, accessed on 10/26/2021). It is shown with the green line in Fig.1,b. We have used the IPO index since 1870 till now. Indeed, along with the AMO, a considerable role in interdecadal climate variability is played by IPO which is characterized by the TPI index (Tripole Index) defined as the difference between the sea surface temperature of the central equatorial Pacific (10°S–10°N, 170°E–90°W) and the mean sea surface temperature of the north-western (25°N–45°N, 140°E– 145°W) and south-western (50°S–15°S, 150°E–160°W) Pacific [62]. The TPI index is shown with the black line in Fig.1,b and located at https://psl.noaa.gov/data/timeseries/IPOTPI/tpi.timeseries.hadisst1.1.data (accessed on 10/23/2020).

The additional index used is the AAO index for the period 1871-2012 (NOAA/NCEP Climate Prediction Center data [63]). The AAO is characterized by the difference of the standardized zonal mean sea level pressures between $40^oS$ and $65^oS$. The AAO index is located at https://psl.noaa.gov/data/20thC_Rean/timeseries/monthly/SAM/sam.20crv2.long.data (accessed on 10/23/2020).

This work presents the analysis of the period since 1880 (the starting date of the temperature data) till 2012 (the ending date of the GHGs forcing and the AAO data). This period includes the data for each of the above variables.

## 3. Method of data analysis

Contributions to temperature trends in each latitudinal zone are estimated on time intervals of the lengths ranging from 5 to 130 years with the aid of AR models analogously to Refs. [29, 30]. The approach fits to the framework of dynamical causal effects [13,61,64] based on the comparison of dynamics under alternative conditions. Namely, each model for a temperature anomaly $T$ accounting for the effects of the GHGs and a natural variability mode $I_M$ is built in the form

$$T_n = a_0 + a_1 T_{n-1} + a_2 I_{GHG,n-1} + a_3 I_{M,n-1} + \xi_n. \qquad (1)$$

Here, $n$ is discrete time (years), $\xi_n$ is noise (residual errors of the model), $I_{GHG}$ is the GHGs radiative forcing, $I_M$ is the index of a climate mode. The AR equation (1) is fitted to the entire observation interval via the ordinary least-squares technique, i.e. via minimization of the sum of squares of residual errors $\xi_n = T_n - a_0 - a_1 T_{n-1} - a_2 I_{GHG,n-1} - a_3 I_{M,n-1}$. The index $I_M$ is either $I_{AMO}$ (the above low-pass filtered AMO index), or $I_{ENSO}$ (ENSO), or $I_{IPO}$ (IPO), or $I_{PDO}$ (PDO), or $I_{AAO}$ (AAO). The estimates of the coefficients $a_0, a_1, a_2, a_3$ are provided with the estimates of their standard errors (deviations) obtained automatically from the same regression estimation under the assumption of delta-correlated finite-variance noise $\xi$. They are computed as the mean squared prediction error multiplied by the inverse of $A^T A$ where A is the matrix of the regressors' values. Significance level at which the null hypothesis of a zero coefficient is rejected can be estimated via the inverse of the standard Gaussian cumulative distribution function evaluated at the estimated value of a coefficient divided by its estimated standard error.

To determine the contributions of the anthropogenic and natural factors to the linear temperature trends for each of the six latitudinal zones over a time interval $[L_{start}, L_{end}]$ with the length $L = L_{end} - L_{start}$, we analyzed time realizations of the AR model (1) with hypothetical regimes for the natural variability modes or the GHGs atmospheric content: instead of the observed time series for a given factor (for example, of a natural variability mode $I_{M,n}$, $n = 1880, ..., 2012$) we "fed" the model with an artificially generated time series $\tilde{I}_{M,n}$ at its input. The initial value of $T$ and the time series of another factor (in the above example, $I_{GHG,n}$, $n = 1880, ..., 2012$) at the model input were the actually observed values. The time series of the "external noise" $\xi_n$ ($n = 1880, ..., 2012$) at the model input was the time series of the residual errors corresponding to the minimum of their sum of squares. The contribution of each factor to the trend was estimated as the difference between the trends of the actually observed values $T_n$ and the model values $\tilde{T}_n$ corresponding to the input signal $\tilde{I}_{M,n}$. This difference of the trends is equal to the linear trend of the temperature difference $\delta T_n = T_n - \tilde{T}_n$. In other words, we assume that the model (1) is equally applicable under the hypothetical condition of an alternative behavior of the factor under study. The trend on each time interval $[L_{start}, L_{end}]$ (with $L$ ranging from 5 to 130 years) was represented by a coefficient $\alpha_{\delta T}$ of the standard linear regression $\delta T_n = \alpha_{\delta T} n + \zeta_n$ obtained via the ordinary least-squares technique. In this way, we estimated the contributions to the temperature trends from the five factors denoting such contributions as $C_{GHG}$ (from GHGs), $C_{AMO}$ (from AMO), $C_{ENSO}$ (from ENSO), $C_{IPO}$ (from IPO), and $C_{AAO}$ (from AAO). We estimated also the actual trend of $T$, i.e. the coefficient $\alpha_T$ in the regression equation $T_n = \alpha_T n + \zeta_n$. To assess the relative role of each factor, we have used the corresponding ratios. Thus, such ratios are $C_{GHG}/\alpha_T$, $C_{AMO}/\alpha_T$, $C_{AMO}/C_{GHG}$, $\tilde{C}_{GHG} = |C_{GHG}|/(|C_{GHG}| + |C_{AMO}|)$, and $\tilde{C}_{AMO} = |C_{AMO}|/(|C_{GHG}| + |C_{AMO}|)$ for the model (1) with GHGs and AMO. Everything is analogous for ENSO, IPO, PDO, and AAO instead of AMO.

This is the simplest method to estimate the contributions to the trends which is based on a minimalistic empirical model. Hence, its advantages are the most reliable statistical estimates and the smallest number of assumptions. Surely, the unit-lag AR model (1) may not be the best one. Further studies with other models deserve efforts. In particular, we have used also linear AR models with greater time lags ranging up to 30 years and with higher AR orders ranging up to five in respect of each variable on the right-hand side of the AR equation. The optimal models turn out to correspond to quite considerable time lags of GHGs and AMO of more than ten years. Their results concerning the estimates of the contributions to the temperature trends are overall similar to those presented here in terms of maximal values of the contributions over different time windows of a given length, while the timing of the maximum contribution (i.e. the location of the respective time window) differs from that for the unit-lag AR model. So, the conclusions about the values of the contributions to the trends are robust to such extension of the model structure. Still, reliability of such larger models deserves separate studies. Accounting for nonlinearity is also potentially significant. So, as a first step, the method in use seems to be the most reasonable among all methods based on the construction and analysis of dynamical models directly from empirical data without any presumed theoretical hypotheses about the climate processes under study.

Concerning the model consistency check which is regarded to be necessary for the attribution [38], all models (1) here agree well with the observed dynamics: The model residuals may be regarded as Gaussian white noise with reasonable accuracy (according to their autocorrelation function and histogram), and the ensembles of model time series obtained under different noise realizations cover the observed temperature time series quite well. Such analysis was done in Refs. [12, 13] for similar AR models of GST with GHGs and solar and volcanic activity (instead of the modes of variability; still, the inclusion of those modes into the models here does not change the above model consistency conclusions). The difference from the consistency check of Ref. [38] is that the noise variance is estimated here from the data for each model separately, so each model agrees with the observed data up to its own variability level. Such flexibility of the model noise parameter is an advantage of the purely empirical approach used here, since the model variability in the attribution studies [38] is specified *a priori* and a model may be claimed there inconsistent only because its variability is too low, while it might be adequate if the variability level were adjusted.

## 4. Results

### 4.1. Empirical AR models

To estimate contributions of different factors to the temperature trends, we have fitted a trivariate AR model (1) with unit time lag for each surface air temperature anomaly $T$ accounting for the impacts of GHGs and one of the natural variability modes. Table 1 presents estimates of the coefficients of the model (1) which characterize sensitivity of the temperature anomalies at different latitudes to changes of the GHGs radiative forcing $I_{GHG}$ and different indices $I_M$ for the entire period under analysis (1880-2012). One can see from those coupling coefficients that the sensitivity to the GHGs changes is smallest for the middle latitudes of SH: it is 2.5 times as small (and so a characteristic time is 2.5 times as large) as that for the tropics of NH. This is explained by the largest thermal inertia for the areas with higher percentage of the sea surface.

*4.1.1. Thermal inertia.* An estimate of a characteristic time of temperature variations is obtained from the dimensionless coefficient $a_1$ which corresponds to a single time step equal here to 1 year. The less the value of $a_1$, the less inertial the process. The closer this coefficient to unity, the greater the relaxation time of the process which equals $\tau \approx 1/(1-a_1)$ years. Naturally, inertia of a latitude band depends on percentage of the sea surface. As mentioned above, the most inertial processes are those in the middle latitudes of SH where the relaxation time of the model (1) is estimated to be about 5 years. Just to compare, the mean relaxation time of the two tropical bands is about 1.7 years (1.8 years for the tropics of SH and 1.5 years for the tropics of NH). Thermal inertia of the middle latitudes of NH is characterized with the relaxation time of 1.4 years in the AR model (1) with AMO. The relaxation time for the polar bands is about 1.3 years. Estimates of the thermal inertia in the AR models (1) with different climatic modes somewhat differ from each other, but the middle latitudes of SH are the most inertial ones in any model.

*4.1.2. Noise intensity estimates.* The variance $\sigma^2$ of the residual errors $\xi_n$ characterizes intensity of the "noise", i.e. external inputs over intra-annual time scales. It is the least in the middle latitudes of SH (0.08 $K^2$), three times as large in both tropical zones, five times as large at the middle latitudes of NH, and 25 times as large in both polar zones being slightly greater in Antarctic.

*4.1.3. Coefficients of the coupling to GHGs.* According to the obtained estimates of the coupling coefficient $a_2$ (Table 1), the impact of GHGs in all latitude bands is significant at the level of $p < 0.05$, i.e. the coefficient estimate exceeds twice its standard error estimate. Most often, the impact of GHGs appears to be far more significant, e.g., even for the middle latitudes of the SH in the model (1) with AAO, it is significant at $p = 0.0005$, i.e. the coefficient estimate is about 3.6 times as great as its standard error. The largest coupling coefficient $a_2 = 0.35\, K/(Wm^{-2})$ is obtained for the Arctic latitudes. It is about twice as small as that for the middle latitudes of the NH and for both tropical zones. It is somewhat greater in the tropics of the SH than in the tropics of the NH. As for the middle and polar latitudes of the SH, this coupling coefficient appears about five times as small as that for the Arctic latitudes, being somewhat less in the middle latitudes than in the polar ones.

Table 1. Estimates of the AR coefficients with 95% intervals (rounded-off to two decimal digits), $\Delta a$ is the doubled standard error of the coefficient estimator. The latitudinal zones are: 1 – Arctic, 2 – middle latitudes of NH, 3 – tropics of NH, 4 – tropics of SH, 5 – middle latitudes of SH, 6 – Antarctic. The numbers in parentheses are the estimates of the significance level at which the hypothesis of zero coefficient is rejected. Bold font shows the coefficients which are nonzero at the level of $p < 0.05$ (highly significant), bold italic – at $0.05 < p < 0.1$ (moderately significant), italic – at $0.1 < p < 0.2$ (weakly significant), normal font shows non-significant values of coefficients.

| Lat. | AR models (1) accounting for | | | | | | | | | |
|---|---|---|---|---|---|---|---|---|---|---|
| | AMO | | ENSO | | IPO | | PDO | | AAO | |
| | $a_2 \pm \Delta a_2$ $K/(Wm^{-2})$ | $a_3 \pm \Delta a_3$ | $a_2 \pm \Delta a_2$ $K/(Wm^{-2})$ | $a_3 \pm \Delta a_3$ | $a_2 \pm \Delta a_2$ $K/(Wm^{-2})$ | $a_3 \pm \Delta a_3$ | $a_2 \pm \Delta a_2$ $K/(Wm^{-2})$ | $a_3 \pm \Delta a_3$ | $a_2 \pm \Delta a_2$ $K/(Wm^{-2})$ | $a_3 \pm \Delta a_3$ $K$ |
| 1 | **0.35±0.10** ($< 10^{-6}$) | **0.70±0.50** (0.006) | **0.30±0.10** ($< 10^{-6}$) | 0.09±0.12 (0.13) | **0.31±0.10** ($< 10^{-6}$) | 0.07±0.11 (0.19) | **0.32±0.10** ($< 10^{-6}$) | 0.03±0.07 (0.52) | **0.32±0.11** ($< 10^{-6}$) | −0.02±0.09 (0.74) |
| 2 | **0.21±0.06** ($< 10^{-6}$) | **0.44±0.25** (0.0005) | **0.16±0.05** ($< 10^{-6}$) | 0.03±0.06 (0.35) | **0.16±0.05** ($< 10^{-6}$) | 0.02±0.05 (0.54) | **0.16±0.06** ($< 10^{-6}$) | 0.01±0.03 (0.73) | **0.15±0.06** ($< 10^{-6}$) | 0.01±0.04 (0.59) |
| 3 | **0.15±0.05** ($< 10^{-6}$) | **0.23±0.20** (0.02) | **0.14±0.05** ($< 10^{-6}$) | *0.04±0.05* (0.16) | **0.14±0.05** ($< 10^{-6}$) | 0.02±0.05 (0.36) | **0.13±0.05** ($< 10^{-6}$) | −0.00±0.03 (0.95) | **0.15±0.05** ($< 10^{-6}$) | ***−0.03±0.03*** (0.06) |
| 4 | **0.17±0.05** ($< 10^{-6}$) | 0.07±0.18 (0.43) | **0.18±0.06** ($< 10^{-6}$) | *0.04±0.05* (0.14) | **0.17±0.06** ($< 10^{-6}$) | 0.02±0.05 (0.46) | **0.16±0.05** ($< 10^{-6}$) | −0.00±0.03 (0.78) | **0.17±0.05** ($< 10^{-6}$) | −0.02±0.03 (0.26) |
| 5 | **0.06±0.03** (0.00006) | 0.00±0.10 (0.96) | **0.06±0.03** (0.00006) | −0.00±0.03 (0.78) | **0.06±0.03** (0.00006) | 0.00±0.02 (0.97) | **0.06±0.03** (0.0001) | 0.01±0.02 (0.45) | **0.05±0.03** (0.0005) | 0.01±0.02 (0.32) |
| 6 | **0.08±0.07** (0.03) | −0.05±0.50 (0.86) | **0.08±0.07** (0.02) | *−0.11±0.12* (0.1) | **0.08±0.07** (0.04) | −0.06±0.12 (0.34) | **0.08±0.07** (0.03) | −0.01±0.07 (0.78) | **0.11±0.09** (0.02) | −0.05±0.10 (0.32) |

*4.1.4. Coefficients of the coupling to AMO.* According to the estimates of the dimensionless coefficient $a_3$ (Table 1), the impact of AMO for the entire period under analysis is essential only in the NH. It is the strongest one at the Arctic latitudes reaching the value of 0.7 significantly nonzero at $p = 0.005$. This value is 1.5 times as small at the middle latitudes of the NH and three times as small at the tropical latitudes of the NH. According to this coefficient, the impact of AMO at the tropical latitudes of SH is 10 times weaker than that in Arctic. The coefficient $a_3$ for the tropics of SH is not significant even at the level of 0.2. So, it is reasonable to consider the contribution of AMO only to the temperature trends in the NH because the relative error of the trend estimate obtained with the model (1) is very close to the relative error of the corresponding coupling coefficient as was also confirmed in our numerical experiments.

*4.1.5. Coefficients of the coupling to ENSO.* According to the estimates of the coefficient $a_3$ (Table 1), the impact of ENSO is strongest at the polar latitudes, its difference from zero is significant at $p = 0.1$ (for Antarctic) and $p = 0.13$ (for Arctic). At Antarctic latitudes, the coupling coefficient to ENSO possesses an opposite sign (negative) in comparison with other latitudes. It is essential that for both tropical zones, this coefficient is two or three times as small. There, it is also less significant with $p = 0.14$ or $0.16$. It is even less and non-significant in the middle latitudes of both hemispheres. Thus, ENSO affects only the tropical and polar zones.

*4.1.6. Coefficients of the coupling to IPO.* According to the estimates of the coefficient $a_3$ (Table 1), the impact of IPO is about twice as small as that of ENSO. The most significant estimate of $a_3$ is obtained for Arctic ($p = 0.19$). Thus, the impact of IPO is most clearly seen in the Arctic zone.

*4.1.7. Coefficients of the coupling to PDO.* The estimates of this coefficient $a_3$ are not significant at all latitudes as reported in Table 1.

*4.1.8. Coefficients of the coupling to AAO.* According to the estimates of the coefficient $a_3$ (Table 1), the impact of AAO is most significant at the tropical latitudes of NH ($p = 0.06$), even more significant than that of ENSO. This coefficient is non-significant at all other latitudes. Still, one can note that it is negative for both tropical zones, where the coupling coefficients from ENSO and IPO are positive. Thus, according to the AAO as a representative of the Pacific Ocean, the latter affects considerably only the temperature variations in the tropical latitudes of the NH.

*4.2. Contributions of various factors to the temperature trends*

Contributions of the GHGs to the trends are estimated in time windows of the lengths ranging from 10 to 130 years. Figure 2 presents those contributions estimated in a moving window of the length of 50 years. To estimate them, the observed values of a temperature are compared to the corresponding model values obtained under the condition of the constant GHGs radiative forcing at the level of 1880 year (see Method). The largest values of the GHGs contribution to the trends are obtained for Arctic latitudes: they equal about 0.2 K/decade and even greater in the last decades. The smallest values are obtained for Antarctic latitudes. For the middle latitudes of NH and tropical latitudes of SH, the maximal values of the GHGs contribution are about 0.2 K/decade, while they are somewhat less for the tropical latitudes of NH and the middle latitudes of SH. Relative values of the GHGs contribution to the trends at various latitudes differ between the time windows.

The contributions of several climatic modes (AMO, ENSO, IPO, PDO, and AAO) are presented in Figure 2 for moving windows of the lengths of 30 (Fig.2,b) and 10 years (Fig. 2,c-g). The largest values are achieved by the contributions to the temperature trends at Arctic latitudes, up to 0.2 K/decade and sometimes even greater. The results of a more detailed analysis for the last decades based on the fixed-end time windows ($L_{end} = 2012$ year) with a moving start point are given in Tables 2-4. Table 2 presents the GHGs contribution to the trends $C_{GHG}$ divided by the angular coefficient of the trend $\alpha_T$ itself (see Method) at different latitudes for the models (1) accounting for the four natural modes. The ratios $C_{GHG}/\alpha_T$ are less than 0.5 only for the relatively short time windows (2-3 decades or shorter) and only at the extratropical latitudes. On the time scales longer than half a century, the GHGs contribution dominates at all latitudes.

Table 3 presents the contributions of the four climatic modes relative to the trends themselves at various latitudes. The ratios $C_{AMO}/\alpha_T$, $C_{ENSO}/\alpha_T$, $C_{IPO}/\alpha_T$, $C_{PDO}/\alpha_T$, and $C_{AAO}/\alpha_T$ do not exceed 0.5 for various time intervals at various latitudes except for Antarctic with non-representative long-term data. The most essential contributions of climatic modes are seen over relatively short time intervals, about 2-3 decades and shorter. Maximal values of the relative contributions are obtained for AMO at various latitudes of NH. A considerable relative contribution of AAO occurs at the middle latitudes of SH, though it is almost statistically insignificant as discussed above (Table 1).

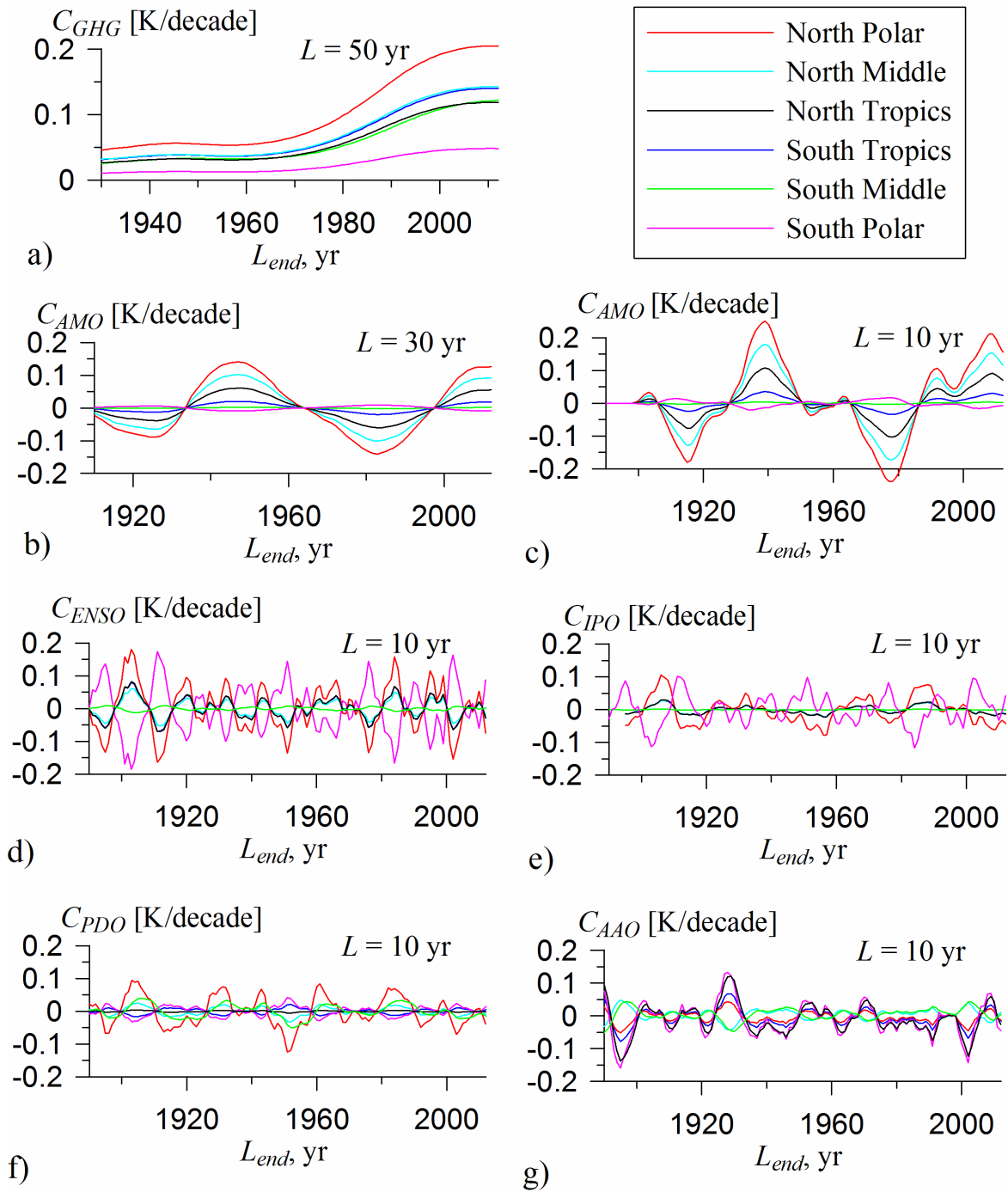

Figure 2. Contributions of different factors to the surface air temperature trends at various latitudes: (a) the contribution of the GHGs estimated in a 50-year moving window with a model (1) accounting for AMO, the estimates with the models accounting for other natural modes are almost indistinguishable; (b) the contribution of AMO in a 30-year moving window; (c) that of AMO in a 10-year moving window; (d) that of ENSO in a 10-year moving window; (e) that of IPO in a 10-year moving window; (f) that of PDO in a 10-year moving window; (g) that of AAO in a 10-year moving window.

Table 2. Estimates of the relative GHGs contributions $C_{GHG}/\alpha_T$. Four numbers in each cell correspond to the models (1) accounting for AMO, ENSO, IPO, PDO, and AAO, respectively. Numbering of the latitudinal zones in the first column is the same as in Table 1. The other four columns show the results for the four different window length with the same window end point $L_{end} = 2012$.

| Lat | 20 years | 30 years | 50 years | 130 years |
|---|---|---|---|---|
| 1 | 0.33, 0.33, 0.34, 0.34, 0.35 | 0.41, 0.42, 0.43, 0.42, 0.43 | 0.64, 0.64, 0.66, 0.64, 0.66 | 0.98, 0.98, 1.01, 0.99, 1.02 |
| 2 | 0.41, 0.41, 0.42, 0.42, 0.40 | 0.44, 0.45, 0.46, 0.45, 0.43 | 0.68, 0.69, 0.70, 0.70, 0.66 | 0.99, 1.00, 1.01, 1.01, 0.96 |
| 3 | 0.61, 0.61, 0.62, 0.61, 0.71 | 0.71, 0.71, 0.72, 0.72, 0.83 | 0.86, 0.86, 0.88, 0.87, 1.00 | 0.98, 0.98, 1.00, 0.99, 1.14 |
| 4 | 1.00, 0.99, 1.01, 0.72, 1.07 | 1.22, 1.21, 1.23, 0.68, 1.32 | 0.95, 0.95, 0.96, 1.07, 1.02 | 1.01, 1.00, 1.02, 1.15, 1.08 |
| 5 | 1.38, 1.38, 1.39, 1.40, 1.26 | 1.86, 1.86, 1.86, 1.88, 1.70 | 1.15, 1.16, 1.16, 1.16, 1.06 | 1.02, 1.02, 1.02, 1.02, 0.93 |
| 6 | -0.52, -0.53, -0.49, -0.52, -0.71 | -0.68, -0.70, -0.65, -0.68, -0.94 | 1.87, 1.93, 1.77, 1.86, 2.58 | 1.38, 1.42, 1.30, 1.37, 1.90 |

Table 3. Estimates of the relative contributions of AMO, ENSO, IPO, PDO, and AAO. Each estimate is reported for the four window lengths (20, 30, 50, and 130 years) with the same window end point $L_{end} = 2012$. Numbering of the latitude zones in the first column is the same as in Table 1. Bold font highlights the values exceeding 0.1 K/decade with the corresponding AR coefficient $a_3$ significant at the level of $p < 0.2$ or at a smaller $p$ (italic shows such values for larger $p$, i.e. with the insignificant AR coefficient $a_3$).

| Lat | $C_{AMO}/\alpha_T$ | | | | $C_{ENSO}/\alpha_T$ | | | | $C_{IPO}/\alpha_T$ | | | | $C_{PDO}/\alpha_T$ | | | | $C_{AAO}/\alpha_T$ | | | |
|---|---|---|---|---|---|---|---|---|---|---|---|---|---|---|---|---|---|---|---|---|
| | 20 | 30 | 50 | 130 | 20 | 30 | 50 | 130 | 20 | 30 | 50 | 130 | 20 | 30 | 50 | 130 | 20 | 30 | 50 | 130 |
| 1 | 0.33 | 0.28 | 0.11 | 0.03 | -0.10 | -0.03 | -0.00 | 0.01 | -0.11 | -0.06 | 0.00 | -0.02 | -0.06 | -0.05 | -0.00 | -0.00 | -0.02 | -0.02 | -0.03 | -0.03 |
| 2 | 0.42 | 0.31 | 0.12 | 0.03 | -0.06 | -0.02 | 0.00 | 0.01 | -0.06 | -0.03 | 0.00 | -0.01 | -0.03 | -0.02 | -0.00 | -0.00 | 0.04 | 0.03 | 0.04 | 0.04 |
| 3 | 0.45 | 0.36 | 0.10 | 0.02 | -0.13 | -0.04 | 0.00 | 0.01 | -0.09 | -0.05 | 0.00 | -0.01 | -0.01 | -0.01 | -0.00 | -0.00 | -0.19 | -0.18 | -0.18 | -0.14 |
| 4 | 0.20 | 0.17 | 0.03 | 0.01 | -0.18 | -0.06 | 0.00 | 0.01 | *-0.12* | *-0.08* | 0.00 | -0.01 | 0.03 | 0.03 | 0.00 | 0.00 | -0.15 | -0.14 | -0.09 | -0.07 |
| 5 | 0.03 | 0.03 | 0.00 | 0.00 | 0.04 | 0.01 | -0.00 | -0.00 | -0.01 | -0.01 | 0.00 | -0.00 | -0.19 | -0.16 | 0.01 | -0.00 | 0.20 | 0.25 | *0.12* | 0.08 |
| 6 | 0.14 | 0.13 | -0.09 | -0.01 | -0.67 | -0.25 | 0.02 | -0.05 | -0.49 | -0.31 | -0.01 | 0.10 | -0.13 | -0.12 | 0.01 | 0.01 | 0.44 | 0.40 | -0.98 | -0.50 |

Table 4 presents the contributions of the four modes relative to the GHGs contribution, i.e. the ratios $\tilde{C}_{AMO} = |C_{AMO}|/(|C_{GHG}| + |C_{AMO}|)$ etc. These ratios are the largest for relatively short time intervals as well. In particular, the values of $\tilde{C}_{AMO}$ are maximal in NH where they reach 0.5 at the middle and polar latitudes for 20-year time intervals. For SH, the values of $\tilde{C}_{AMO}$ are considerably smaller. The trend contributions of ENSO are most significant not only at the tropical latitudes, but also at the polar ones. At the polar latitudes, considerable trend contributions of IPO are also revealed. Considerable trend contributions of AAO in SH are seen not only on the time scales of about two or three decades, but also on the scale of half a century, though their statistical significance is negligible as discussed above (Table 1).

Table 4. Contributions of the natural variability modes relative to the GHGs contribution $\tilde{C}_{AMO}$, $\tilde{C}_{ENSO}$, $\tilde{C}_{IPO}$, and $\tilde{C}_{AAO}$. Each value is given for the four different window lengths with the same window end point $L_{end} = 2012$. Numbering of the latitude zones in the first column is the same as in Table 1. Bold font highlights the values exceeding 0.1 with the corresponding AR coefficient $a_3$ significant at the level of $p < 0.2$ or at a smaller $p$ (italic shows such values for larger $p$, i.e. with the insignificant AR coefficient $a_3$).

| Lat | AMO | | | | ENSO | | | | IPO | | | | PDO | | | | AAO | | | |
|---|---|---|---|---|---|---|---|---|---|---|---|---|---|---|---|---|---|---|---|---|
| | 20 | 30 | 50 | 130 | 20 | 30 | 50 | 130 | 20 | 30 | 50 | 130 | 20 | 30 | 50 | 130 | 20 | 30 | 50 | 130 |
| 1 | **0.50** | **0.40** | **0.11** | 0.03 | *0.23* | 0.08 | 0.00 | 0.01 | **0.24** | **0.13** | 0.00 | 0.02 | *0.14* | *0.11* | 0.00 | 0.00 | 0.06 | 0.04 | 0.04 | 0.03 |
| 2 | **0.51** | **0.41** | **0.12** | 0.03 | *0.12* | 0.04 | 0.00 | 0.01 | *0.11* | 0.07 | 0.00 | 0.01 | 0.06 | 0.04 | 0.00 | 0.00 | 0.09 | 0.07 | 0.06 | 0.04 |
| 3 | **0.42** | **0.34** | **0.10** | 0.02 | *0.18* | 0.06 | 0.00 | 0.01 | *0.13* | 0.07 | 0.00 | 0.01 | 0.01 | 0.01 | 0.00 | 0.00 | **0.21** | **0.18** | **0.15** | **0.11** |
| 4 | *0.17* | *0.12* | 0.03 | 0.01 | *0.16* | 0.05 | 0.00 | 0.01 | *0.11* | 0.06 | 0.00 | 0.01 | 0.04 | 0.04 | 0.00 | 0.00 | *0.12* | *0.10* | 0.08 | 0.06 |
| 5 | 0.02 | 0.02 | 0.00 | 0.00 | 0.03 | 0.01 | 0.00 | 0.00 | 0.01 | 0.00 | 0.00 | 0.00 | *0.12* | 0.08 | 0.00 | 0.00 | *0.14* | *0.13* | 0.10 | 0.07 |
| 6 | *0.22* | *0.16* | 0.05 | 0.01 | *0.56* | *0.27* | 0.01 | 0.03 | *0.50* | *0.32* | 0.01 | 0.07 | *0.20* | *0.15* | *0.01* | *0.01* | *0.38* | *0.30* | *0.28* | *0.21* |

## 5. Conclusions

According to the presented results, the contributions of the key modes of natural variability to the surface air temperature trends on relatively short time intervals within three decades reaches and can exceed (in absolute value) ± 0.2 K/decade, while they are not considerable as compared to the contribution of the GHGs atmospheric content rise on time intervals about half a century and longer. The GHGs contribution always increases, reaching 0.2 K/decade in the last two decades and even somewhat exceeding this value. The GHGs contribution dominates on time intervals of about half a century and longer, and sometimes even on shorter intervals.

These estimates are especially important to compare the contemporary trends of the surface air temperatures at various latitudes of NH and SH with assessment of the relative roles of the key natural variability modes on different temporal horizons. In particular, an overall increase of the Antarctic sea ice extent in the last decades (up to the last years) according to the satellite data (available only starting from 1970s) accompanied with the global warming and fast decrease of the Arctic sea ice extent is related to the general decrease of the surface temperature at sub-Antarctic latitudes which occurred since 1970s till 2016 when a fast decrease of the sea ice extent in Southern Ocean was noticed. This is related to regional manifestations of the natural climate oscillations with periods up to several decades accompanied with the global century-scale warming and a relatively weak temperature trend over the ocean in SH. The cross-correlation and the cross-wavelet analyses evidence a significant coherence and a negative correlation between the surface temperature and the sea ice extent in the last decades both in Arctic and Antarctic (see e.g. [33]). According to the estimates obtained in this work, a contribution of AAO, in particular, to the surface temperature trends at different latitudes of SH and tropical latitudes of NH is seen not only on the time scales of 20-30 years, but also on the time scales of half a century and longer. It should be taken into account in constructing future projections of regional climate changes on the basis of climate models. It is necessary for such models to describe adequately the natural climate variability and its contribution to the regional temperature trends on various temporal horizons. Currently, many important modes of climate variability are reproduced by such models, but there are also modes of variability that are not simulated well enough (e.g. [34]).

This study was carried out under the support of the Russian Science Foundation (grant No. 19-17-00240) with the use of the results on peculiarities of climate variability in the middle and high latitudes of Northern Hemisphere obtained under the support of the Ministry of Science and Higher Education of Russian Federation (Agreement No. 075-15-2020-776).